\documentclass[twocolumn,showpacs,preprintnumbers,amsmath,amssymb,superscriptaddress,float]{revtex4}

\usepackage{graphicx}
\usepackage{dcolumn}
\usepackage{bm}
\usepackage{txfonts}

\begin{document}

\title{Spectrum bandwidth narrowing of Thomson scattering X-rays with energy chirped electron beams from laser wakefield acceleration}

\author{Tong Xu}
\affiliation{Key Laboratory for Laser Plasmas (Ministry of Education), Department of Physics and Astronomy, Shanghai Jiao Tong University, Shanghai 200240, China}
\author{Min Chen}
\email{minchen@sjtu.edu.cn} \affiliation{Key Laboratory for Laser
Plasmas (Ministry of Education), Department of Physics and
Astronomy, Shanghai Jiao Tong University, Shanghai 200240, China}
\author{Fei-Yu Li}
\affiliation{Key Laboratory for Laser Plasmas (Ministry of Education), Department of Physics and Astronomy, Shanghai Jiao Tong University, Shanghai 200240, China}
\author{Lu-Le Yu}
\affiliation{Key Laboratory for Laser Plasmas (Ministry of Education), Department of Physics and Astronomy, Shanghai Jiao Tong University, Shanghai 200240, China}
\author{Zheng-Ming Sheng}
\email{zmsheng@sjtu.edu.cn} \affiliation{Key Laboratory for Laser
Plasmas (Ministry of Education), Department of Physics and
Astronomy, Shanghai Jiao Tong University, Shanghai 200240, China}
\affiliation{SUPA, Department of Physics, University of
Strathclyde, Glasgow G4 0NG, UK}
\author{Jie Zhang} \affiliation{Key
Laboratory for Laser Plasmas (Ministry of Education), Department
of Physics and Astronomy, Shanghai Jiao Tong University, Shanghai
200240, China} \affiliation{Beijing National Laboratory of
Condensed Matter Physics, Institute of Physics, CAS, Beijing
100190, China}

\begin{abstract}
We study incoherent Thomson scattering between an ultrashort laser
pulse and an electron beam accelerated from a laser wakefield. The
energy chirp effects of the accelerated electron beam on the final
radiation spectrum bandwidth are investigated. It is found that
the scattered X-ray radiation has the minimum spectrum width and
highest intensity as electrons are accelerated up to around the
dephasing point. Furthermore, it is proposed  the electron
acceleration process inside the wakefield can be studied by use of
$90^\circ$ Thomson scattering. The dephasing position and beam
energy chirp can be deduced from the intensity and bandwidth of
the scattered radiation.

\end{abstract}


\maketitle
\date{\today}
Nowadays high bright X-ray sources based on intense laser pulses
have attracted a lot of attentions due to their wide applications,
relative low cost, easy operation, and unique
characters~\cite{Corde2013}. Compared with traditional accelerator
based synchrotron radiation sources these new sources have shorter
duration and easier synchronism with lasers, which makes them more
flexible for pump-probe technique. Many mechanisms are proposed to
generate X-ray radiations. For incoherent sources, radiation based
on inner shell electron excitation through laser-solid interaction
are extensively studied~\cite{ChenLM2008,Levy2010,Kugland2008};
electron betatron radiation inside a laser wakefield or radiation
from laser Thomson scattering through laser gas interactions are
also studied~\cite{Kneip2010,Cipiccia2011,Schnell2013,ChenS2013}.
For coherent sources, radiation based on high harmonics generation
in laser-solid interaction or coherent Thomson scattering from
laser nanometer electron sheet interaction are
studied~\cite{Shaw2013,Baeva2006,Wu2011}. The latter has attracted
more and more interests recently due to the feasibility of
electron sheet generation resulting from laser plasma
interactions~\cite{Li2013,Wu2010,Liu2013}.

In this paper, we focus on incoherent radiation from laser
electron Thomson scattering. We use the electron beam accelerated
from a laser wakefield accelerator~\cite{Tajima1979,Esarey2009},
which may allow easy synchronization between the electron beams
and laser pulses. We show that, due to the energy chirp of the
electrons accelerated inside a wakefield, the final spectrum of
the scattered radiation can be narrowed compared with a normal un-chirped
electron beam. In the meanwhile, we propose to deduce the
acceleration process inside a wakefield  by diagnosing the
radiation spectrum. This provides a new possible approach to
detect the wakefield.

For simplicity, here we use a one-dimensional (1D) model for the
electron beam accelerated in a wakefield. This gives a
proof-of-principle description for Thomson scattering with chirped
electron beams.  As one knows for the electrons accelerated along
the same phase space trajectory in a wakefield whose normalized
phase velocity is $\beta_p=v_p/c$, the relation between its longitudinal
momentum ($p_z$) and phase position $\psi$ satisfies:
$p_z=\beta_p\gamma_p^2[H+\phi(\psi)]\pm\gamma_p\{\gamma_p^2[H+\phi(\psi)]^2-\gamma_\perp^2\}^{1/2}$,
where $H$ is the Hamiltonian along the specific trajectory in the
phase space set by the wakefield, $\phi$ is the potential of the wake and
$\gamma_p=(1-\beta_p^2)^{-1/2}$. Usually electrons injected at
different instant time can get different acceleration length which
contributes to the final energy spread of the accelerated electron bunch.
Besides this, electrons injected at the same physical position may
be put into different trajectories in the phase space, which also
results in energy spread to the final beam~\cite{Chen2012}. Thus the
whole beam usually shows energy chirp before accelerating to the
dephasing length where the fastest electrons begin to decelerate.
To get small energy spread one usually let the electrons be
accelerated further a little bit over the dephasing position where
both acceleration and deceleration happen to the fast and slow
electrons, respectively. At this point, the electrons have
relative high energy in the beam center and low energy both in
front and at end. Prior to and after this stage, beams show
monotonic energy chirp. As shown in the following, the Thomson
scattering from these electrons is quite different and the
spectrum actually can be used as a diagnostic method to find the
dephasing position.

On the other hand, laser Thomson scattering is a wellknown process
in which electrons oscillate inside  a laser field and radiate new
electromagnetic field~\cite{Esarey2002}. Normally classical
electrodynamics is enough to describe such process once the
emitted photon's energy in the electron rest frame is far less
than the electron energy, i.e. quantum recoil effect can be
neglected. The radiation spectrum shows synchrotron radiation
characters. Depends on the laser intensity the radiation may
include high harmonic components or show single peak character.
The normalized laser vector potential ($a=eA_{\perp}/m_ec^2$)
acts as the strength parameter of an undulator or wiggler inside a
synchrotron facility. The resonant frequency is
$\omega_n=4\gamma_{z0}^2n\omega_L/(1+a^2/2+\gamma_{z0}^2\theta^2)$.
Here $\omega_L$ is the laser frequency and $\theta$ is the
radiation angle related to the longitudinal motion direction of
the electron. For values of $a\ll 1$, the laser pulse acts as an
undulator and emitted radiation by a single electron will be
narrowly peaked about the fundamental resonant frequency
$\omega_1$(n=1). As $a$ increases, the pulse is more like a
wiggler and the emitted radiation will appear at harmonics of the
resonant frequency as well ($\omega_n$). The final spectrum
consists of many closely spaced harmonics. For electrons with
different energy ($\gamma_{z0}$) inside a bunch these spectrum
incoherently superimposed and the final spectrum appears
broadband. A continuum of radiation is generated which extends out
to a critical frequency, $\omega_c$, beyond which the radiation
intensity goes down.

The resonant frequency along the axis ($\theta=0$) is
$\omega_1=4\gamma_{z0}^2\omega_L/(1+a^2/2)$.  Even for a
monoenergetic electron bunch (with same $\gamma_{z0}$) the
radiation spectrum will be broadened due to the different laser
intensities ($a$) along the interaction path. Usually for a
normal temporally Gaussian pulse [$a\propto
\exp(-\zeta^2/c^2T^2)$] with $\zeta=z-ct$, the maximum intensity
is in the center of the pulse. If one use an electron beam with
special energy chirp to make
$\gamma_{z0}^2(\zeta)/[1+a^2(\zeta)]={\rm {constant}}$, the emitted
radiation will be narrowed. We call such a beam a matched beam
with the laser pulse. As discussed above, for the electrons
accelerated to the dephasing position inside the laser wakefield,
they can automatically match with a normal pulse. Recently
Ghebregziabher \textit{et al.} has proposed another way to reduce
the spectrum width by using chirped laser pulse. In their scheme
once the laser frequency and intensity satisfy
$\omega_L(\zeta)/[1+a^2(\zeta)/2]={\rm {constant}}$, the emitted
resonant frequency on axis will be the same during the scattering
process with a $180^\circ$ laser-beam interaction geometry, in
which a monoenergetic beam is assumed~\cite{Ghebregziabher2013}.
These two schemes (electron energy matching or laser frequency
matching) share similar ideas. The main difference is that in our
scheme a normal laser pulse and a natural accelerated electron
beam accelerated from a wake field can be used and the $180^\circ$
interaction geometry is not necessary.

\begin{figure}
  \includegraphics[width=8cm]{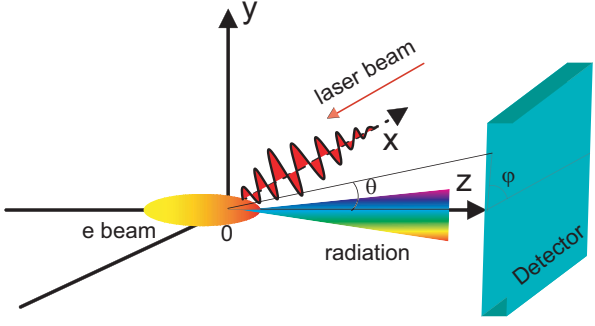}
\caption{Schematic view of Thomson scattering. The laser pulse is
focused at zero point and propagates along the -x direction.  The
electron beam propagates along z direction.} \label{FigSetup}
\end{figure}


We use the VDSR code to simulate the electron laser Thomson
scattering process~\cite{Chen2013}. The classical radiation
calculation model  is used and end-points effects are
considered~\cite{Chen2013}. In the simulation we fix the laser
pulse and vary the electron beam properties. The spatial and
spectrum distribution of the far field incoherent radiation is
recorded by a virtual detector inside the code. The incoherent
radiation is calculated by summing the radiation intensity
contributed by each single electron.
\begin{figure}
  \includegraphics[width=7.5cm]{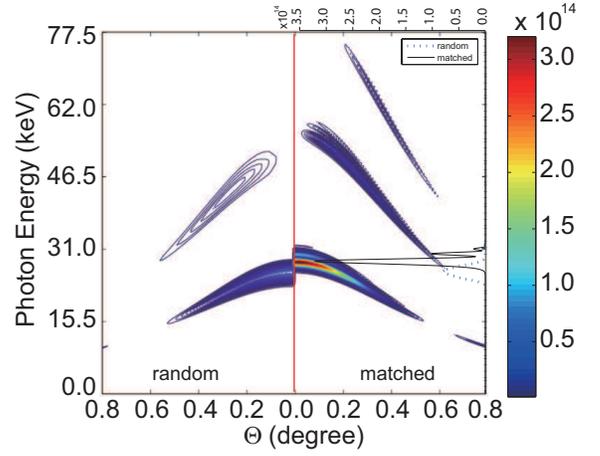}
\caption{Angular and energy distribution of the radiations for the
random unmatched (left part) and  matched (right part) beams
interaction with a laser pulse. The on axis (along z direction)
radiation spectra are shown by the solid curve for the matched
beam and by the dashed curve for the unmatched beam.}
\label{normalthomson}
\end{figure}

We first study the $90^\circ$ Thomson scattering from a laser
pulse interaction with a normal un-chirped electron beam.  A linearly
polarized laser pulse with normalized electric field of
$a=eE/m\omega_0 c \propto a_0\exp(-r^2/W^2-t^2/T^2)$ propagates
along the $-x$ direction with $a_0=0.5$. The pulse is set to be
focused at (x,y,z)=(0,0,0).  The pulse center is initially at
(60,0,0) with the normalization length of $\lambda_0$. Here
$T=20T_0$, $W=20\lambda_0$, $\omega_0=2\pi/T_0$ and $T_0=2.67\rm
fs$, $\lambda_0=0.8\mu m$ are the period and wavelength of the
laser, respectively. An electron beam with charge of 10 pC is
launched from (0,0,-80). The beam has a cylindrical shape with
transverse radius of $0.1\lambda_0$ and longitudinal size of
$40.0\lambda_0$. The initial longitudinal momenta of the electrons
are sampled as $p_z=\langle p_z\rangle (1+\alpha\cdot\delta p_z)$
with $\alpha$ a random number uniformly distributed between
[-1,1], central momentum $\langle p_z\rangle=98m_e c$, and
momentum spread $\delta p_z=0.02$. It corresponds to a beam with
central energy of 50 MeV and rms energy spread of $1.1\%$. For
simplicity the beam initially has zero transverse emittance
($\delta p_\perp=0$). From above parameters we know the centers of
the electron beam and laser beam will collide at zero point. Due to
Doppler effect, the main radiation from the electron beam will be
focused within a cone with half open angle of $1/\gamma$
\textit{rad} which is around $0.58^\circ$. In our code the
detector records the radiation intensity $d^2I/d\omega d\Omega$
which is a function of $\omega$, $\theta$, and $\varphi$. According
to the scattering parameters, we set the collection angle as:
$0^\circ<\theta<0.8^\circ$ with totally 80 bins and
$0<\omega<5.0\times10^4\omega_0$ with totally 5000 bins. We fix
$\varphi=0^\circ$ here. The setup of the laser beam interaction is
shown in Fig.~\ref{FigSetup}.

\begin{figure}
  \includegraphics[width=8cm]{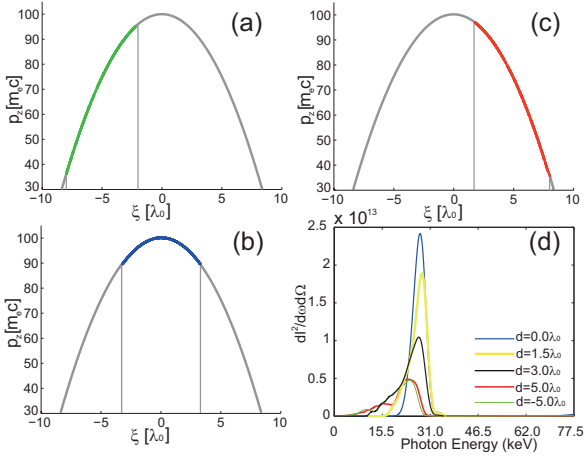}
\caption{Electron beam distribution in the phase space. $\xi=z-v_pt$ is the spatial coordinates of the electrons in the wake rest frame. Here $v_p$ is the phase speed of the wake, which is close to the group velocity of the driver pulse $v_g\simeq c\sqrt{1-n_e/n_c}$ inside a plasma with density of $n_e$. $n_c=m_e\omega_0^2/4\pi e^2$ is the critical density for the driver pulse. Electrons between the two vertical lines and along the trajectory are used for the Thomson scattering calculation. Here the slice energy spread of the beam is set to be 0 in the simulations. (a-c) correspond to beams with central position $-5\lambda_0$, $0$, and $5\lambda_0$ away from the dephasing position, respectively. (d) shows the on-axis radiation spectra from the electron beams with different central position deviations from the dephasing position.}
\label{wake1}
\end{figure}

The final angular and energy distributions of the radiation for
this unmatched beam is shown in the left part of
Fig.~\ref{normalthomson}. The on axis ($\theta=0^\circ$) radiation
spectrum for this case is also shown in this figure by the right
dashed curve. As we see the radiation shows harmonics characters
and the first resonant radiation frequency is around
$2\gamma^2/(1+a_0^2/2)\omega_0\approx 1.7\times 10^4\omega_0$. It
corresponds to a photon energy of 26.46 keV which is close to
the simulation result of 25.93 keV. The FWHM spread of the emitted
radiation spectrum is about 3.12 keV.

The radiation spectrum can be optimized if we use a matched
chirped electron beam.  To find the matching condition, we
calculate the laser field felt by each single electron. Since the
laser pulse in our interaction is not relativistic but the
electrons are relativistic, we assume the electrons have ballistic
trajectories when they go through the laser. We only discuss the
condition where the interaction angle between the beam and the
pulse is $90^\circ$. For other interaction angles, the optimization process is similar.
We assume a single electron is initially at
$(0,0,z_0+\xi_0)$ and moves along z direction with a speed close to
the light speed $c$. Here $z_0$ is the initial beam center
position and $\xi_0$ is the electron's position relative to the
beam center. The center of the beam and the pulse collide with
each other at the zero point at time $t=0$. It is easy to see the
electrons with coordinate of $\xi_0$ will feel the laser field as
$a=a_0\exp[-(\xi_0+ct)^2/W^2-t^2/T^2]$. To make a matched
condition, one should make the variation of
$\gamma^2(\xi_0)/(1+a^2/2)$ as small as possible for all the
electrons with different initial positions ($\xi_0$). To satisfy
this in a second simulation the beam's longitudinal momentum is
sampled as $p_z=p_{zmax}
\sqrt{[1+a_0^2\exp(-2\xi^2/W^2)/2]/(1+a_0^2/2)}$ with
$p_{zmax}=100~m_e c$. The laser and electron parameters except
this longitudinal momentum distribution are the same as the
previous simulation (shown in Fig.~\ref{normalthomson} left part). In this case, it corresponds to a beam with
maximum energy of 51.1 MeV and rms energy spread of $1.1\%$.  The
final radiation spectrum is shown in the right part of
Fig.~\ref{normalthomson}. The on-axis radiation spectrum is shown
by the solid curve. It shows the radiation peak locates at photon
energy of 28.33 keV and the FWHM spread of the radiation spectrum
is 1.19 keV. As we can see the radiation spectrum has been
narrowed to $38\%$ of the previous spectrum width and the on axis
radiation intensity has been increased more than 4.1 times. These
simple calculations show the importance of a matched beam on the
final X-ray radiations.

It should be pointed out that the above optimized scheme needs a
very precise synchronization  between the laser pulse and the
electron beam with an approximately tolerable delay deviation as
large as the beam length. For separated laser pulse and beam this
is obviously quite difficult. However for electron beam accelerated
by laser plasma wakefield it is not a big issue since the
accelerated beam is usually just behind the driver laser
with a delay length of the wake wavelength. The scattered laser
pulse can be split from the driver pulse or easily synchronized
with such pulses. On the other hand as we mentioned before, for
the electron beam inside a wakefield, the energy chirp always
exists. In the following we show the Thomson scattering of a laser
wakefield accelerated electron beam with different energy chirps.

\begin{figure}
  \includegraphics[width=8cm]{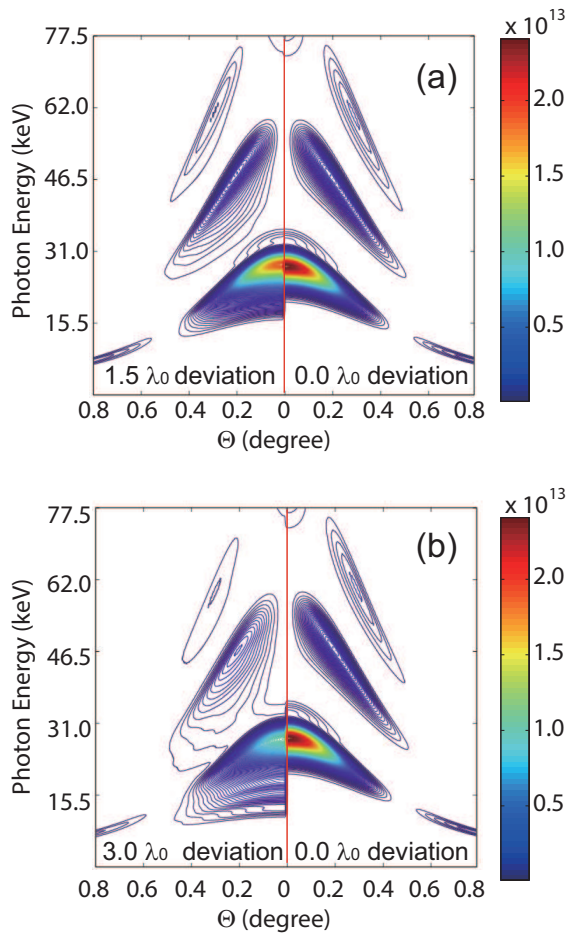}
\caption{(a) Angular and energy distribution of the radiations for a beam with central position deviation of $1.5\lambda_0$ (left part) and a matched (right part) beam interaction with a laser pulse. (b) corresponds to a beam with central position deviation of $3.0\lambda_0$ case.}
\label{wake2}
\end{figure}

To compared with the above simulation results, in the simulations
the same laser parameters are used  and a trajectory in the phase
space with maximum accelerated longitudinal momentum of $100 m_ec$
is used. Typical electron distributions at three different
acceleration distances are used to scatter. The distributions of
electrons in the phase space are shown in Fig.~\ref{wake1}(a-c).
The first beam [see Fig.~\ref{wake1}(a)] whose center in the phase
space is about $5\lambda_0$ in front of the dephasing point has a
negative energy chirp. The second one [see Fig.~\ref{wake1}(b)] whose center just locates at
the dephasing position has highest energy in
the center part. The third one [see Fig.~\ref{wake1}(c)] is
symmetric with the first one in the phase space and it has a
positive energy chirp. It deserves to point out that the distances
labeled here are the coordinate values ($\delta\xi$) in the rest
frame of the laser wakefield. The real distance between the
acceleration positions is approximately $\delta\xi/(1-v_p/c)\simeq
(2n_c/n_e)\delta\xi$. For a typical plasma density of
$n_e=10^{19}/\rm cm^3$, a distance of $5\lambda_0$ in phase space
corresponds to a real distance of $2.1 \rm mm$ which is far larger
than a typical focusing spot size of the transverse scattering
pulse (usually $\sim10\mu \rm m$ level). As we will show the second
beam is somewhat matched with the laser pulse and gives best
radiation spectrum among the three cases. The radiated spectra are
shown in Fig.~\ref{wake1}(d). As one can see when the electrons
are around the dephasing position, the radiated spectrum has the
highest intensity and narrowest bandwidth. Using beams beside this
position, radiations have lower peak intensity and the emitted
radiation shows lower frequency. In our cases more than 4.9 times
higher radiation has been obtained by using the matched electron
beam compared with the beam with $5\lambda_0$ deviation from the
dephasing position and the bandwidth has been reduce almost by
half (from 7.58 keV to 4.58 keV). As we can see the negative chirped
beam gives similar radiations as the positive chirped beam. This is
because in our cases the laser electron interactions for this two
cases are exactly symmetric if the laser is a Gaussian beam and
the two beams collision with each other at their own
central peaks. We also checked the effect of other beam center
deviation lengths related to the dephasing point in the phase
space on the spectrum. The radiations of deviation of
$1.5\lambda_0$ and $3.0\lambda_0$ are shown in Fig.~\ref{wake1}(d)
by the yellow and black lines, respectively. As we can see the
more deviation from the dephasing position, the wider the
radiation spectrum and lower radiation intensity. By studying the
radiation spectrum one may deduce the acceleration process inside
the wakefield such as the position of the dephasing point.

The radiation angular distributions for beams with deviation
length of $1.5\lambda_0$ and  $3.0\lambda_0$ are shown in
Fig.~\ref{wake2}. Again one can see the radiation is also more
focused in space and narrowed in spectra for a matched
electron beam. The more deviation from the dephasing point the
wider the radiation spectrum.

In summary we studied the Thomson scattering from laser
interaction with energy chirped electron beams accelerated from
laser wakefield acceleration. By matching the beam energy
distribution with the laser pulse intensity, a narrowed spectrum
of radiation is obtained. The maximum radiation intensity is about
5 times larger than an unmatched beam. Our scheme maybe used to
optimize the Thomson/Compton scattering spectrum in future
experiments for laser plasma based X-ray sources~\cite{ChenS2013}.
By using $90^{\circ}$ Thomson scattering, this
method can also be used to detect the electron dephasing position
in the LWFA. Radiation from the electrons through betatron
radiation inside the wake or scattering from other magnetic or
laser undulator has already been used to characterize the electron
beams~\cite{Corde2012,Plateau2012}. For other scattering angles
(other than $90^\circ$), to narrow the radiation spectrum one can
use an electron beam with transverse spatial energy chirp or use a
laser pulse with transverse spatial chirp. Such laser pulse has
already been used to improve the HHG from laser gas
interaction~\cite{Kim2013hhg}. We should mention that the slice energy
spread for electron beams inside the phase space has been
neglected in our model. This energy spread is due to the injection
of the electrons into different trajectories inside the wakefield.
Numerous computational studies have proved that usually this
spread is far less than the one due to different acceleration
length, which corresponds to the energy spread included here. In
our simulation we also found once the slice energy spread
contribution is less than 10 percent of total energy spread, the
spectrum variation results from the mechanism described here is
obvious.

This work is supported in part by the National Basic Research
Program of China (Grant No. 2013CBA01504), the National Science
Foundation of China (Grant No. 11205101, 11121504, 11374209, and
11374210), and the MOST international collaboration project
0S2013GR0050. MC appreciates supports from Shanghai Science and
Technology Commission (Grant No. 13PJ1403600) and National 1000
Youth Talent Project of China. We also appreciate helpful
discussions with Dr. Feng Liu. The simulations were carried out on
the $\Pi$ supercomputer in Shanghai Jiao Tong University.


\end{document}